\documentstyle[aps]{revtex}
\newcommand{\dir}{FIGS}

\input epsf

\begin{document}

\title{Interfacial profiles between coexisting phases in thin films:\\
Cahn Hilliard treatment versus capillary waves
}

\author{
Kurt Binder, Marcus M\"{u}ller, Friederike  Schmid, and Andreas Werner
\\
{\small Institut f{\"u}r Physik, WA 331, Johannes Gutenberg Universit{\"a}t}
\\
{\small D-55099 Mainz, Germany}
}
\date{\today, submitted to J.Stat.Phys.}

\maketitle

\begin{abstract}
A symmetric binary mixture (A,B) below its critical temperature $T_c$ of unmixing is considered in a thin
film geometry confined between two parallel walls, where it is assumed that one wall prefers $A$ and the other wall prefers $B$.
Then an interface between the coexisting unmixed phases is stabilized, which (above the wetting transition temperature) occurs
in the center of the film for an average concentration of $c=1/2$.
The problem is considered how the concentration profile $c(z)$ across the thin film depends on the film thickness $D$. By Monte Carlo 
simulation of a lattice model for a polymer mixture it is shown
that for relatively small $D$ the width of the interface scales like 
$w \propto D$ while for larger $D$ a crossover to a behavior $w \propto \sqrt{D}$ occurs.
This behavior is explained by phenomenological theories: it is shown that the behavior at small $D$ can be
understood by a suitable extension  of the Cahn Hilliard ``gradient-square''-type theory, while the behavior for
large $D$ can be traced back to the behavior of capillary waves exposed to a short range potential by the walls.
Corrections due to fast concentration variations, as they occur in the strong segregation limit
of a polymer mixture, can be accounted for by self-consistent field theory. Subtle problems occur, however, with respect to the
proper combination of these theories with the capillary wave approximation, particularly at intermediate values of $D$.
\end{abstract}

\section{Introduction}

The theoretical understanding of interfacial profiles between coexisting
phases has been a longstanding challenge \cite{1,2,3,4,5,6,7,8,9,10,11,12,13,14,15,16,17,18,19,20,21,22,23,24,25,26}. After Cahn and
Hilliard \cite{6} in a seminal pioneering work proposed the ''gradient
square'' theory for binary mixtures, to derive the (intrinsic \cite{1})
interfacial profile in the framework of a mean field type treatment, this
approach has been extended to treat interfaces in polymer solutions \cite{7}%
, polymer blends \cite{8,9,10,10,11,12,13,16,17,18,19,20,21,22,23,25,26} block copolymer mesophases and
other modulated phases in complex fluids \cite{14,15,24} etc. Of course, for
a full understanding of the interfacial profile one has to amend the mean
field treatment by a consideration of statistical fluctuations neglected in
this treatment, notably long wavelength fluctuations of the local position
of the center of the interfacial profile (usually termed ''capillary waves'' 
\cite{1,2,3,4,19,21,25,26,27,28,29,30,31}).

Here we focus on another extension of the ''gradient square'' theory, namely
the profile of an interface in a geometry confined between two parallel
walls a distance D apart. While interfaces bound to external walls have been
considered for a long time in the context of wetting transitions \cite
{2,3,23,32,33,34,35,36,37,38}, interfaces confined by two walls have been studied only more
recently \cite{39,40,41,42,43,44,45,46}, focusing on the interface localization -
delocalization transition that may occur in this geometry. In the present
study, however, we investigate another aspect of these confined interfaces
that has been discovered in recent simulations\cite{47,48,49} and experiments
on polymer mixtures \cite{47,50}: there occurs a significant reduction of
the intrinsic width $w_{0}$ of the interfacial profile already for
relatively thick films, $D\gg w_{0}$. We expect that this phenomenon is
important also for a description of interface-location - delocalization
transitions and wetting phenomena, since it implies a renormalization of
parameters entering the effective interface Hamiltonian\cite{51}. This
''squeezing'' of the intrinsic interfacial profile also is important for
experimental studies that try to deduce effective interaction parameters from
measured concentration profiles \cite{52} of polymer mixtures, but apply the
standard theory for interfacial widths of strongly segregated systems \cite
{8,9,13,16,23} that ignores the finite thickness D of the thin film that is
used.

The outline of the present paper, hence, is as follows: we first (Sec. 2)
present a phenomenological theory of this squeezing of interfacial profiles
in thin films, by a fairly straightforward extension of the theory of Parry
and Evans\cite{42}, who did not pay attention to this effect. As these
authors, we consider only short range forces of the walls for simplicity,
although for a quantitative description of experiments \cite{47,50,52,53}
long range van der Waals forces should be used (numerical evidence shows
that for long range forces there even is a stronger squeezing of the
interfacial profile due to confinement than for short range forces \cite{49})
. We show that there is a rather broad range of D where $w_{0}$ varies
nearly linear with D. While this treatment implicitly implies the case of a
weakly segregated mixture, where concentration variations are slow, we then
(Sec. 3) treat the alternative case of a strongly segregated polymer
mixture where the profile varies across the interface from $\phi _{A}=0$ to 
$\phi _{A}=1$, but nevertheless a ''square gradient'' theory applies, in the
limit of long chain length of the polymers\cite{8,9}. The case of
intermediate segregation is more complicated, since even a free, unconfined
interface of a polymer mixture exhibits a profile involving two different
lengths \cite{12}. We study such intermediate cases in Sec. 4 by a numerical
version of the self-consistent field theory \cite{8,9,11,13,16,17,18} that uses
Monte Carlo - generated single chain configurations as an input \cite
{21,51,54}. Sec. 5 then summarizes and discusses our results, paying
attention also to the problem how the present results can be combined with
the effects of capillary waves in this confined geometry (which for $%
D\rightarrow \infty $ lead to an apparent \cite{47,48} interfacial width $%
w\propto \sqrt{D}$ while the intrinsic width $w_{0}$ saturates at a finite
value -- the mean field result for an unconfined free interface).

\section{Reduction of intrinsic interfacial widths of coexisting symmetrical
mixtures confined by ''competing'' walls}

We consider here a symmetrical binary mixture (AB) sufficiently close to the
bulk critical point such that the concentration variations of interest are
of rather long range, and a Ginzburg-Landau type description in terms of the
order parameter $m(z)=\left\{ \left[ c(z)-c_{\rm crit}\right] /c_{\rm crit}\right\} $%
(here c(z) is the concentration of species A at position z and the critical
concentration c$_{\rm crit}$ of species A is c$_{\rm crit}$ $=1/2$) is applicable.
We consider a thin film geometry with two ''competing'' planar walls,
oriented perpendicular to the $z$-direction, a distance $D$ apart (Fig.\ 1).
''Competing'' walls mean that the left wall preferentially attracts species
B, and the right wall preferentially attracts A, and thus the concentration
c(z), as well as m(z), increases monotonically from left to right. We take
these forces between the walls and the molecules (or atoms,
respectively) of the mixture of short range, i. e. they act locally on
particles adjacent to the walls only, for the sake of simplicity, and
furthermore we assume their absolute strength equal, so that the profiles
m(z) shown in Fig. 1 must be antisymmetric around the point z=0, m(z=0)=0
i.e. the center of the profile in the middle of the thin film. Of course,
here we have also assumed that these forces exerted on the particles on the
wall are sufficiently strong, so that no symmetry breaking due to the
localization of the interface is possible at the considered temperature.
Note that for large enough D the interface localization-delocalization
transition would occur close to the wetting transition temperature of the
corresponding semi-infinite system\cite{39,40,41,42,43,44,45,46}, and recall the argument due
to Cahn\cite{32} that close to the bulk critical point of a mixture the
surfaces always should be wet.

We now emphasize that in general we expect five distinct regions in m(z)
given the assumptions made above: In the first region close to the left
wall, m(z) will decay from its value $-m_{0}(D)$ right at the wall to the
value $-m_{b}$, characteristic for the left branch of the coexistence curve, 
$c_{\rm coex}^{(1)}$, of the binary mixture in the bulk $\left\{ -m_{b}=\left[
c_{\rm coex}^{(1)}-c_{\rm crit}\right] /c_{\rm crit},~+m_{b}=\left[
c_{\rm coex}^{(2)}-c_{\rm crit}\right] /c_{\rm crit}\right\} $. This decay occurs over a
length scale of the same order as the correlation length $\xi _{b}$ in the
bulk. In the second regime where $m(z)\approx -m_{b}$ \{centered at $-z_{b}$
defined by $m(-z_{b})=-m_{b}$\} the slope of the profile dm(z)/dz reaches a
minimum, and in the limit D$\rightarrow \infty $ the profile would become
completely flat. Near the center of the film we have the interfacial profile
in a strict sense, which for large D is described simply by \cite{1,2,3,4,5,6} $%
m(z)=m_{b}\tanh \left[ z/w_{0}(D)\right] $ with $w_{0}(D\rightarrow \infty
)=2\xi _{b}$. For not so large D, however, the ''intrinsic width'' w$_{0}(D)$
of this profile is significantly reduced, cf. Fig. 1 (lower part), and this
interfacial regime is no longer so clearly distinguished from the first
regime, since the second regime where m(z) stays nearly flat when m(z)$%
\approx -m_{b}$, nearly has disappeared \{and so does the fourth regime
where m(z)$\approx +m_{b}$ near z=z$_{b}$\}. In the fifth regime we have the
increase from m(z)$\approx +m_{b}$ to $m(z=D/2)=m_{0}(D)$, which clearly is
dependent on the precise nature of the boundary conditions at the walls
(unlike $m_{b}$ and $\xi _{b}$ which are not dependent on these boundary
conditions, of course.

From this (qualitative) expectation that the concentration profile exhibits
five distinct regimes (the central regime is expected to have a width of 2$%
\xi _{b}$, the others must have at least a width of $\xi _{b}$ in order for
this picture to make sense) one already can predict that the asymptotic
regime of this interfacial behavior is only reached for $D\gg 6\xi _{b}$,
while a rather nontrivial behavior \{including a significant reduction of
the width $w_{0}(D)$ in comparison to $2\xi _{b}$\} is expected if this
condition is not met. Of course, Fig. 1 makes sense only for the case of
rather weak segregation, close enough to the bulk critical point, where $%
m_{b}$=1 for our normalization). If one can reach a case of strong
segregation but still is in a regime of complete wetting, then $%
m_{b}\rightarrow 1$ and also $m_{0}\left( D\right) =1,$ and one obtains a
much simpler situation than shown in Fig. 1, with $m(z)$ varying from $-1$
at z=$-D/2$ to $+1$ at $z=D/2$ in a single step \{a profile varying similar
to $\tanh \left[ z/w_{0}(D)\right] $\}. It has been predicted \cite{55} that
such a situation (strong segregation in the bulk but complete wetting of a
surface) is typically realized in polymer mixtures, and this prediction was
confirmed by corresponding Monte Carlo simulations \cite{48,56}. We shall
return to this case of strong segregation in the next section.

In order to calculate the explicit form of this nontrivial profile in Fig.\
1 from a Ginzburg-Landau type theory \cite{1,2,3,4,5,6,42}, it is convenient to
rescale the order parameter $m(z)$ by its bulk value, $M(Z)\equiv m(z)/m_{b}$%
, and also lengths are measured in units of the bulk correlation length, $%
Z=z/\xi_{b}$. Then the rescaled free energy of the film can be written as
follows\cite{3,42}

\begin{eqnarray}
\xi _{b}^{-1}F(D)&=&\int\limits_{-D/(2\xi _{b})}^{+D/(2\xi _{b})}{\rm d}Z\; \left[ 
-\frac{1}{2}M^{2}(Z)+\frac{1}{4}M^{4}(Z)+(dM/dZ)^{2}\right]  \nonumber \\
&&+\left( \xi _{b}/\lambda \right) \left[ M^{2}(-D/2)+M^2(D/2)\right] +H_{1}\left[
M(-D/2)-M(D/2)\right] 
\label{gl1}
\end{eqnarray}

Note that at $z=+D/2$ we have applied a (normalized) field $H_{1}$ and at $%
z=-D/2$ a (normalized) field $- H_{1}$ coupling to the order parameter; this
term represents the preferential attraction of species A to the right wall,
species B to the left wall. As is well known\cite{3}, one must allow in the
''bare'' wall free energy \{the second line of Eq.(\ref{gl1})\} a term proportional to
$M^{2}$ as well (to account for the effect of ''missing neighbors''
and possible change of interactions near the wall, etc.). Following common
notation\cite{3}, the coefficient of this term has been put inversely
proportional to the so called ''extrapolation length'' $\lambda $ which is a
microscopic length (of the order of the interaction range in a model with
short range interactions). Note that F(D) is normalized per unit surface
area of the walls, and in units where the bulk free energy per unit volume
is $-1/4$.

Now the profile M(Z) in Fig. 1 results from seeking the minimum of the free
energy functional, Eq.(\ref{gl1}). The Euler-Lagrange equation corresponding
to Eq.(\ref{gl1}) is 
\begin{equation}
-M(Z)+M^{3}(Z)-2\frac{{\rm d}^{2}M(Z)}{{\rm d}Z^{2}}=0  \label{gl2}
\end{equation}
with the boundary conditions
\begin{equation}
dM/dZ-(\xi _{b}/\lambda )M=H_{1}/2,\qquad Z=-D/2\quad ,  \label{gl3}
\end{equation}

\begin{equation}
dM/dZ+(\xi _{b}/\lambda )M=H_{1}/2,\qquad Z=+D/2\quad .  \label{gl4}
\end{equation}

After multiplication of the Euler-Lagrange equation with $M^{^{\prime
}}\equiv {\rm d}M/{\rm d}Z$ one can integrate once to find 
\begin{equation}
\left( \frac{dM}{dZ}\right) ^{2}=-\frac{1}{2}M^{2}(Z)+\frac{1}{4}M^{4}(Z)+%
\frac{1}{4}-\Delta p\left( D\right) =\left[ M^{2}\left( Z\right) -1\right]
^{2}/4-\Delta p\left( D\right) \quad ,  \label{gl5}
\end{equation}

where the integration constant was denoted as $1/4-\Delta p\left( D\right) ,$
for the sake of consistency of notation with the work of Parry and Evans 
\cite{42}. For $D\rightarrow \infty $, we expect that $dM/dZ\rightarrow 0$
for $M(Z)=\pm 1$ (corresponding to the flat regions 2 and 4 in Fig. 1) and
thus $\Delta p(D\rightarrow \infty )\rightarrow 0$.

Eq.(\ref{gl5}) can also be combined with the boundary conditions, Eqs.(\ref {gl3}), (\ref{gl4}), and hence we conclude, using $\tilde{M}=M(Z=D/2\xi _{b})$
as an abbreviation,

\begin{equation}
(\tilde{M}\xi _{b}/\lambda -H_{1}/2)^{2}=(\tilde{M}-1)^{2}/4-\Delta p(D),
\label{gl6}
\end{equation}

and from Eq.(\ref{gl5}) we find, invoking the symmetry $M(-Z)=-M(Z),$

\begin{equation}
D/2\xi _{b}=\int\limits_{0}^{\tilde{M}}dM/\sqrt{(M^{2}-1)^{2}/4-\Delta p}.
\label{gl7}
\end{equation}

Using once more Eq.(\ref{gl6}) one finds a closed expression for $\tilde{M}$
as function of D in terms of the inverse function expressed as a quadrature,

\begin{equation}
D/2\xi _{b}=\int\limits_{0}^{\tilde{M}}dM/\sqrt{\left[ \left( M^{2}-1\right)
^{2}-(\tilde{M}^{2}-1)^{2}\right] /4+(\tilde{M}\xi _{b}/\lambda
-H_{1}/2)^{2}.}  \label{gl8}
\end{equation}

The maximum slope of M(Z) at Z=0 \{where also M(Z=0)=0, see Fig. 1\} can be
written as

\begin{equation}
\frac{dM}{dZ}\left|_{\rm max} =\left[ 1/4-\Delta p(D)\right] ^{1/2}=\left[ (\tilde{M}%
\xi _{b}/\lambda -H_{1}/2)^{2}-\tilde{M}^{4}/4+\tilde{M}^{2}/2\right]
^{1/2}.\right.  \label{gl9}
\end{equation}

Noting that for $D\rightarrow \infty $ we must have $\Delta p=0$ (otherwise
the integral in Eq.(\ref{gl7}) would converge to a finite constant) we can
obtain from Eq.(\ref{gl6}) straightforwardly the local order parameter $%
\tilde{M}_{\infty }$ at the surface, for this situation of complete wetting,

\begin{equation}
\tilde{M}_{\infty }\xi _{b}/\lambda -H_{1}/2=-(\tilde{M}_{\infty }^{2}-1)/2.
\label{gl10}
\end{equation}

The (second-order) wetting transition occurs when \cite{55} $\tilde{M}%
_{\infty }\rightarrow 1,$ i. e. for H$_{1c}=2\xi _{b}/\lambda $. Hence the
situation shown in Fig. 1 requires $H_{1}>H_{1c}$ and then $\tilde{M}%
_{\infty }>1$. In general, we expect for finite D that $\tilde{M}<\tilde{M}%
_{\infty }$ and for D$\rightarrow \infty $ we have a smooth convergence of $%
\tilde{M}$ towards $\tilde{M}_{\infty }$. This expectation can be verified
from Eq.(\ref{gl6}) by writing $\tilde{M}=\tilde{M}_{\infty }-\delta M$ and
expanding to first order in $\delta M$, using also Eq.(\ref{gl10}),

\begin{equation}
\left[ 2\left( \xi _{b}/\lambda \right) \left( \tilde{M}_{\infty }\xi
_{b}/\lambda -H_{1}/2\right) +\tilde{M}_{\infty }(\tilde{M}_{\infty
}^{2}-1)\right] \delta M=\Delta p,  \label{gl11}
\end{equation}
or
\begin{equation}
\delta M=\left[ (-\xi _{b}/\lambda +\tilde{M}_{\infty })(\tilde{M}_{\infty
}^{2}-1)\right] ^{-1}\Delta p.  \label{gl12}
\end{equation}
For large $D$ we shall find that $\Delta p$ is very small $\{ \Delta p \propto \exp(-D/2\xi_b)$,
see Eq.(\ref{gl18}) below$\}$, and thus $\tilde{M}\approx\tilde{M}_\infty$.
For D$\rightarrow 0$ we see from Eq.(\ref{gl7})
that then also $\tilde{M}%
\rightarrow 0$, and hence in this limit the integral can be evaluated
expanding the square root as
\begin{equation}
\left( \frac{M^{4}}{4}-\frac{M^{2}}{2}+\frac{1}{4}-\Delta p\right)
^{-1/2} \approx \left( \frac{1}{4}-\Delta p\right) ^{-1/2}\left[ 1+M^{2}/(1-4\Delta
p)-+....\right] ,  \label{gl13}
\end{equation}
and hence we obtain
\begin{equation}
D/2\xi _{b}=\left[ \tilde{M}/\left( \frac{1}{4}-\Delta p\right)
^{1/2}\right] \left[ 1+\tilde{M}^{2}/(3-12\Delta p)-+....\right] .
\label{gl14}
\end{equation}

In this limit where D/2$\xi _{b}\approx \tilde{M}/\left( \frac{1}{4}-\Delta
p\right) ^{1/2}\ll  1$ we also have $M(Z)=2\tilde{M}Z\xi _{b}/D$, the profile
is simply linear, and from Eq.(\ref{gl9}) we conclude that
\begin{equation}
\frac{dM}{dZ}\left|_{\rm max} =(1/4-\Delta p)\right. ^{1/2}\approx H_{1}/2,\qquad
D \rightarrow 0.  \label{gl15}
\end{equation}

We expect that Eq.(\ref{gl15}) remains true as long as $\tilde{M} \ll 1$, i. e. $%
D/2\xi _{b} \ll 2/\sqrt{\frac{1}{4}-\Delta p}\approx 4/H_{1}$. Note that in our
normalization H$_{1}$ is dimensionless and typically larger than one, since $%
H_{1}>H_{1c}=2\xi _{b}/\lambda \gg 1$. The fact that the inverse of $%
dM/dZ|_{\max }$, which can be considered as a measure of the (normalized)
interfacial width $w_{0}(D)$, remains non-vanishing as $D\rightarrow 0$ can
be understood from the fact that the gradient energy $(dM/dZ)^{2}$ in Eq.\
(\ref{gl1}) disallows too steep gradients and hence for $D\rightarrow 0$ one
can no longer have a profile from $-m_{b}$ to $+m_b$ in Fig. 1 but only from $-\tilde{m%
}$ to $+\tilde{m}$ where also $\tilde{m}\equiv \tilde{M}m_{b}\rightarrow 0$
as D$\rightarrow 0$, because otherwise $(dM/dZ)^{2}$ would diverge in this
limit.

We now consider the inverse limit of Eq.(\ref{gl7}), namely $D\rightarrow
\infty $ when $\tilde{M}\rightarrow \tilde{M}_{\infty }$. This means that
the profile in Fig. 1 develops a very broad plateau near $m(z)\approx m_{b}$
for $z>0$ \{and m(z) $\approx -m_{b}$ for z%
\mbox{$<$}%
0, respectively\}. As a consequence, the dominating part of the integral in
Eq.(\ref{gl7}) comes from $M\approx 1$, and this suggests to approximate the
integral as follows
\begin{eqnarray}
\frac{D}{2\xi_b} &=& \int_0^1 {\rm d}M/\sqrt{(M^2-1)^2/4-\Delta p} +  \int_1^{\tilde{M}} {\rm d}M/\sqrt{(M^2-1)^2/4-\Delta p} \nonumber \\
		 &\approx&  \int_0^1 {\rm d}M/\sqrt{(M^2-1)^2/4-\Delta p} +  \int_1^{\tilde{M}} {\rm d}M/\sqrt{(M-1)^2-\Delta p} \nonumber \\
		 &=&  \int_0^1 {\rm d}M/\sqrt{(M^2-1)^2/4-\Delta p} + \ln \left( \frac{\tilde{M}-1+\sqrt{(\tilde{M}-1)^2-\Delta p}}{\sqrt{-\Delta p}}\right) \label{gl16}
\end{eqnarray}

The first integral in the last line of Eq.(\ref{gl16}) is independent of the boundary
condition, and tends to $\ln \left[ C/\sqrt{-\Delta p}\right] $ for $%
\Delta p\rightarrow 0,$where a numerical evaluation shows that the
constant C is roughly $C\approx 4.$\ The second term in the last line of
Eq.(\ref{gl16}) becomes in this limit $\ln\left[2(\tilde{M}-1)/\sqrt{-\Delta p}\right]$, and hence we conclude
\begin{equation}
\frac{D}{2\xi_b} \approx \ln \left[ \frac{2C(\tilde{M}-1)}{-\Delta p}\right] \label{gl17}
\end{equation}
or
\begin{equation}
-\Delta p = 2 C (\tilde{M}-1)\exp(-D/2\xi_b). \label{gl18}
\end{equation}
Hence Eq.(\ref{gl9}) yields in this limit for the maximum slope of the profile at $%
Z=0 $
\begin{equation}
\frac{{\rm d}M}{{\rm d}Z}\Big|_{\rm max} = \sqrt{1/4+2C(\tilde{M}-1)\exp(-D/2\xi_b)} = \frac{1}{2}\sqrt{1+8C(\tilde{M}-1)\exp(-D/2\xi_b)} \label{gl19}
\end{equation}
From Eq.(\ref{gl19}) we obtain our central result for the width $w_{0}\left( D\right)$
in Fig. 1, namely
\begin{equation}
\frac{w_0(D)}{w_0(\infty)} = \sqrt{1+8C(\tilde{M}-1)\exp(-D/2\xi_b)}, \qquad D \gg 2\xi_b. \label{gl20}
\end{equation}
Eqs.(\ref{gl19}) and (\ref{gl20}) describe the approach to the limit of an ''intrinsic'' interface
in the bulk, which has the width $w_{0}\equiv w_0\left( \infty \right) =2\xi
_{b}$ in our units. Since $\tilde{M}$  can appreciably exceed one, and the constant $%
8C\approx 32$ is rather large, the approach of $w_{0}\left( D\right) $ to $%
w_{0}\left( \infty \right) $ is rather slow, and hence a significant reduction of
the interfacial width of a confined interface is predicted. Since in the
considered limit we may replace $\tilde{M}$ by $\tilde{M}_\infty$ in Eq.(\ref{gl20}), we can use Eq.(\ref{gl10}) to write,
for $H_{1}$near $H_{1c},$ using also $C\approx 4$,
\begin{equation}
\frac{w_0(D)}{w_0(\infty)} = \sqrt{1+16(H_1-H_{1c})\exp(-D/2\xi_b)/(1+H_{1c}/2)}. \label{gl21}
\end{equation}

In Fig.2 the approximation, Eq.(\ref{gl20}), is compared to a full evaluation of Eqs. (\ref{gl8}) and (\ref{gl9})
which can only be done numerically: for a given choice of $\xi
_{b}/\lambda $ and $H_{1}$ we first obtain $\tilde{M}_\infty$ from Eq.(\ref{gl10}). Then we take for
this choice of $\xi _{b}/\lambda $ and $H_{1}$ several values of $\tilde{M}<\tilde{M}_\infty$ and
evaluate the values of $D/2\xi _{b}$ that then correspond to each $\tilde{M}$. Since $%
dM/dZ|_{\rm max}$ is readily given in terms of $\tilde{M}$  as well, from Eq.(\ref{gl9}), one can easily
plot the ratio $w_0(D)/w_0(\infty) = dM/dZ|_{{\rm max},D=\infty} / dM/dZ|_{{\rm max},D}$ versus $D/2\xi_b$.
From Fig.2 we see that for large $D/\xi _{b}$ the reduction of $w_{0}\left(
D\right) /w_{0}\left( \infty \right) $ is only dependent on $\tilde{M}_\infty$, as expected
from Eq.(\ref{gl20}) since in this limit $\tilde{M}\approx\tilde{M}_\infty$ can be used there. In addition, for $%
2\lesssim D/\xi _{b}\lesssim 8$ the variation of $w_{0}\left( D\right) $
with $D$ is approximately linear, and a saturation of $w_{0}\left( D\right) $
at $w_{0}\left( \infty \right) $ only occurs for $D/\xi _{b}\geq 12$ ,
as predicted on the basis of the qualitative discussion of Fig.\ 1 at the
beginning of this section.

If for a given $\tilde{M}$ the corresponding $D$ has been evaluated from Eq.(\ref{gl8}), $\Delta
p\left( D\right) $ being known from Eq.(\ref{gl6}), one can use Eq.(\ref{gl5}) to compute the
full profile $M\left( Z\right) $ in analogy with Eq.(\ref{gl7}), namely
\begin{equation}
Z = \int_0^{M(Z)} {\rm d}M'/\sqrt{(M'^2-1)^2/4-\Delta p}  \label{gl22}
\end{equation}

Putting $M\left( Z\right) =1$ here one obtains the value $Z_{b}=z_{b}/\xi
_{b}$ defined in Fig.1. All the details of Fig. 1 thus can be verified
explicitly.

\section{Self-consistent field treatment for confined interfaces of strongly
segregated polymer mixtures}

The Monte Carlo simulations of a lattice model of a strongly segregated
polymer mixture between competing walls\cite{48} already have provided numerical
evidence (Fig. 3) that also in this limit there is a regime where the
observed interfacial width $w$ apparently scales as $w\propto D,$ despite
the fact that interfacial fluctuations yield a strong increase of $w$ with $%
D $ for $D\gg w_{0}\left( \infty \right) $ as well.

In order to present a theoretical understanding for this problem of
interface squeezing by confining walls in strongly segregated mixtures, we
present here an extension of the theory due to Helfand {\em et al.}\cite{8,9} The basic
quantities are the probability densities $q_{A}\left( z,t\right) $, $%
q_{B}\left( z,t\right) $ that one end of an $A$ or $B$ chain with degree of
polymerization $t$ is at $z.$ The position of the other chain end is
arbitrary. For a blend of $A$ and $B$ homopolymers of the same chain length $%
N_{A}=N_{B}=N,$ these functions satisfy modified diffusion equations,
\begin{equation}
\frac{\partial}{\partial t} q_A(z,t) = \frac{b^2}{6}  \frac{\partial^2 q_A(z,t)}{\partial z^2} - w_Aq_A(z,t), \label{gl23}
\end{equation}
\begin{equation}
\frac{\partial}{\partial t} q_B(z,t) = \frac{b^2}{6}  \frac{\partial^2 q_B(z,t)}{\partial z^2} - w_Bq_B(z,t), \label{gl24}
\end{equation}
where the ''fields'' $w_{A},w_{B}$ derive as $w_{\alpha}=\partial f/\partial
\rho_\alpha$ from a free energy density that depends on the densities $\rho_A(z)$, $\rho_B(z)$
\begin{equation}
\rho_\alpha(z) = \frac{1}{N} \int_0^N {\rm d}t q_\alpha(z,N-t) q_\alpha(z,t), \qquad \alpha=A,B  , \label{gl25}
\end{equation}
as
\begin{equation}
f = \chi\rho_A\rho_B + \frac{\zeta}{2}(\rho_A+\rho_B-1)^2. \label{gl26}
\end{equation}
Here the factor $\left( k_{B}T\right) ^{-1}$ is absorbed in the free energy
density, $\chi$ is the Flory-Huggins parameter that causes the unmixing of the
polymer mixture ($\chi N \gg 1$ defines the strong segregation limit), and $\zeta$ controls
the inverse compressibility of the blend.

While Eqs.(\ref{gl23}) - (\ref{gl26}) define the general framework of the self consistent field
theory of polymer blends, we here are only interested in the limit $%
N\rightarrow \infty $ and treat also the blend as incompressible. As a
consequence of the limit $N\rightarrow \infty $, one can take\cite{8,9} $\rho_\alpha = q_\alpha^2$
$(\alpha = A,B)$ and put $\partial q_\alpha/\partial t=0.$ Redefining the
units of length such that $b/\sqrt{6\chi}\equiv 1$, Eqs.(\ref{gl23}) and (\ref{gl24}) then can be replaced by
\begin{equation}
\frac{{\rm d}^2q_\alpha(z)}{{\rm d}z^2} - \frac{1}{2} \frac{\partial f}{\partial q_\alpha} = 0, \qquad \alpha=A,B. \label{gl27}
\end{equation}

We can interpret Eqs.(\ref{gl27}) again as Euler-Lagrange equations of the Lagrangian
\begin{equation}
{\cal L} = \int {\rm d}z \left\{ \frac{1}{2} f(q_A,q_B) + \frac{1}{2} \left(\frac{{\rm d}q_A}{{\rm d}z}\right)^2 + \frac{1}{2} \left(\frac{{\rm d}q_B}{{\rm d}z}\right)^2 \right\}, \label{gl28}
\end{equation}
where we have a ''conservation law'' for the Hamiltonian
\begin{equation}
{ H} = \frac{1}{2}  \left(\frac{{\rm d}q_A}{{\rm d}z}\right)^2 + \frac{1}{2} \left(\frac{{\rm d}q_B}{{\rm d}z}\right)^2 - \frac{1}{2} f(q_A,q_B), \label{gl29}
\end{equation}
if $d/dz$ is reinterpreted as a ''time'' derivative, as usual.

In the inhomogeneous case the quantity of interest is the excess free
energy ( again units of $(k_BT)^{-1}$ being used )
\begin{equation}
F_{\rm exc} = \int_{z_0}^{z_1} {\rm d}z (f -w_A\rho_A - w_B\rho_B) \label{gl30}
\end{equation}
which after integrating by parts can be rewritten as
\begin{equation}
F_{\rm exc} = -2{ H}(z_1-z_0) + 2\int_{z_0}^{z_1} {\rm d}z \left[ \left(\frac{{\rm d}q_A}{{\rm d}z}\right)^2 + \left(\frac{{\rm d}q_B}{{\rm d}z}\right)^2 \right]
- \left[ q_A \frac{{\rm d}q_A}{{\rm d}z}\right] \Big|_{z_0}^{z_1}
- \left[ q_B \frac{{\rm d}q_B}{{\rm d}z}\right] \Big|_{z_0}^{z_1} \label{gl31}
\end{equation}

We now simplify the problem by requiring strictly local incompressibility $\rho_A+\rho_B = q_A^2+q_B^2=1$ 
everywhere in the system. It is convenient to express this condition
using polar coordinates
\begin{equation}
q_A = \sin\phi, \qquad q_B= \cos\phi, \label{gl32}
\end{equation}
which amounts to replace the Lagrangian in Eq.(\ref{gl28}) by
\begin{equation}
{\cal L} = \int {\rm d}z \left\{\frac{1}{2} \left( \frac{{\rm d}\phi}{{\rm d}z} \right)^2 +\frac{1}{2} \tilde{f}(\phi) \right\}, \label{gl33}
\end{equation}
where
\begin{equation}
\tilde{f}(\phi) = \rho_A\rho_B = \frac{1}{4}\sin^2(2\phi). \label{gl34}
\end{equation}
Then the excess free energy formulated in Eqs.(\ref{gl30}) and (\ref{gl31}) becomes
\begin{equation}
F_{\rm exc} = -(z_1-z_0)\Delta / 4 + 2 \int_{z_0}^{z_1} {\rm d}z \left( \frac{{\rm d}\phi}{{\rm d}z} \right)^2 \label{gl35}
\end{equation}
where $\Delta $ is a constant of motion resulting as
\begin{equation}
\Delta/8 = \frac{1}{2} \left( \frac{{\rm d}\phi}{{\rm d}z} \right)^2  - \frac{1}{2} \tilde{f}(\phi)  \label{gl36}
\end{equation}
This constant $\Delta$ is the analog of the constant $-\Delta p(D)$ of the previous section.

As a first step, we apply this formalism to an interface in an infinitely
thick system without any boundary effects, such that $\lim_{z \to \pm \infty} dq_\alpha/dz=0$, and also $\lim_{z \to \pm \infty} f(q_A,q_B)=0$.
Since thus the constant of motion $\{$Eq.(\ref{gl29})$\}$ ${ H}=0$ and also $\Delta=0$
, we have
\begin{equation}
F_{\rm exc} = 2 \int_{-\infty}^{+\infty} {\rm d}z \left( \frac{{\rm d}\phi}{{\rm d}z} \right)^2 = 2 \int_0^{\pi/2} {\rm d}\phi/ \frac{{\rm d}z}{{\rm d}\phi}, \label{gl37}
\end{equation}
and the corresponding Euler-Lagrange equation
\begin{equation}
\frac{{\rm d}\phi}{{\rm d}z} = \frac{1}{2} \sin 2\phi \label{gl38}
\end{equation}
is solved by
\begin{equation}
\ln(\tan \phi) = z, \qquad \phi = {\rm arctan}(\exp(z)) \label{gl39}
\end{equation}
and noting Eqs.(\ref{gl32}) this is recognized as the familiar $\frac{1}{2}\left[
1+ {\rm tanh} \left( \frac{z}{2}\right) \right] $ profile for the density,\cite{8}
\begin{equation}
q_A/q_B = \exp(z), \qquad q_A = \frac{\exp(z)}{1+\exp(z)} \label{gl40}
\end{equation}
From Eqs.(\ref{gl37}) and (\ref{gl38}) one notes that $F_{\rm exc}=1$ is the free energy cost of the
interface in our units.

Next we consider a semi-infinite system, in order to discuss wetting behavior
in the strong segregation limit. Analogously to Eq.(\ref{gl1}) , we choose a bare
surface free energy of the form
\begin{equation}
F_{\rm surf}^{\rm bare} = \frac{2}{\lambda} (\rho_A^0-\rho_B^0)^2 - h_1(\rho_A^0-\rho_B^0) \label{gl41}
\end{equation}
where $\rho_A^0,\rho_B^0$ are the densities at the surface, and $h_{1}$ is a ''surface
field'' in suitable units, $\lambda$ being the analog of the ''extrapolation
length'' used in Eq.(\ref{gl1}).

We now note that the above profile, Eqs. (\ref{gl39}) and (\ref{gl40}) is cutoff by the surface\cite{23,55},
and thus it is convenient to introduce the angle $\alpha=\pi-2\phi_0$ that
corresponds to the values $\rho_A^0,\rho_B^0$reached in the surface plane. The bulk part
of the excess free energy can be written as (note $\Delta =0$ still holds)
\begin{equation}
F_{\rm exc}^{\rm bulk} = \int_0^\infty {\rm d}z \left( \frac{{\rm d}\phi}{{\rm d}z} \right) = \int_0^{\pi/2-\alpha/2} {\rm d} \phi \;\; \sin2\phi = \frac{1}{2}\left(1+\cos \alpha\right) \label{gl42}
\end{equation}
and combining Eqs. (\ref{gl41} and (\ref{gl42}) the total excess free energy becomes
\begin{equation}
F_{\rm exc}^{\rm tot} = \frac{1}{2} + \cos\alpha \left(\frac{1}{2}-h_1\right) + \cos^2 \alpha/\lambda . \label{gl43}
\end{equation}
This free energy takes a minimum for
\begin{equation}
\cos \alpha = \frac{\lambda}{2} \left(h_1-\frac{1}{2}\right).  \label{gl44}
\end{equation}

A second order wetting transition occurs for $\lambda >0,$ at
\begin{equation}
h_1^c = \frac{1}{2} + \frac{2}{\lambda} \label{gl45}
\end{equation}
while for $1/\lambda \leq 0$ one has a first order wetting (as is
actually observed in the simulation\cite{57} of the model shown in Fig. 3).

Finally we are now in the position to consider the case of thin films of
finite thickness $D$ , where now the constant of motion $\Delta $ is
nonzero. The quantity that we wish to calculate is $w_{0}\left( D\right) $,
defined from the inverse slope in the center of the profile
\begin{equation}
w_0(D) = \left(2 {\rm d}\rho_A/{\rm d}z \right)^{-1}\Big|_{z=0} = \left(2 {\rm d}\phi/{\rm d}z \right)^{-1}\Big|_{z=0} = \frac{1}{\sqrt{1+\Delta}} \label{gl46}
\end{equation}
where we have used that
\begin{equation}
\frac{{\rm d}\rho_A}{{\rm d}z} = \frac{{\rm d}q_A^2}{{\rm d}z} = 2 q_A \frac{{\rm d}q_A}{{\rm d}z} = sin (2\phi) \frac{{\rm d}\phi}{{\rm d}z} \label{gl47}
\end{equation}
and we note that in the center of the profile $\phi =\pi/4$ since there
$\rho_A=\rho_B$ by symmetry. From Eqs.(\ref{gl34}) and (\ref{gl36}) we have deduced that
\begin{equation}
\left(\frac{{\rm d}\phi}{{\rm d}z} \right)^2 = \left[ \Delta/4+\tilde{f}(\phi))\right] = \left[ \Delta + \sin^2(2\phi)\right]/4, \label{gl48}
\end{equation}
and thus the counterpart of Eq.(\ref{gl22}) for the profile becomes
\begin{eqnarray}
z &=& \int {\rm d}\phi\;/\left(\frac{{\rm d}\phi}{{\rm d}z}\right) = 2 \int_{\pi/4}^{{\phi}} {\rm d}\tilde{\phi}\;/\sqrt{\Delta+\sin^2(2\tilde{\phi}) } \nonumber \\
&& = \frac{1}{\sqrt{1+\Delta}} E_1\left(\pi/2-\alpha,\frac{1}{\sqrt{1+\Delta}}\right)  \qquad {\rm if} \qquad \Delta>0, \label{gl49} \\
&& = E_1\left(\pi/2-{\rm arccos}\left(\frac{\cos \alpha}{\sqrt{1+\Delta}}\right),\sqrt{1+\Delta}\right),  \qquad {\rm if} \qquad \Delta<0, \label{gl50}
\end{eqnarray}
$E_{1}$ denoting the elliptic integral of the first kind. Denoting the angle
$\phi(z=0)=\phi_0$, and correspondingly $\alpha=\pi-2\phi_0$, the excess free energy can then be written
as
\begin{eqnarray}
F_{\rm exc}^{\rm bulk} &=& - \frac{1}{4}\Delta D + 4\int_{\pi/4}^{\phi_0}  {\rm d}\phi\;\frac{{\rm d}\phi}{{\rm d}z}  \nonumber \\
		       &=& - \frac{1}{4}\Delta D + 2 \int_{\pi/4}^{\phi_0} {\rm d}\tilde{\phi} \sqrt{\Delta + \sin^2(2\tilde{\phi})}  \nonumber \\
		       &=& \left\{ \begin{array}{ll}
			   -\frac{1}{4} \Delta D + \sqrt{1+\Delta} \;E_2\left(\pi/2-\alpha,\frac{1}{\sqrt{1+\Delta}}\right) & {\rm for} \qquad \Delta>0 \\
			   -\frac{1}{4} \Delta D + E_2\left(\pi/2-{\rm arccos}\left(\frac{\cos \alpha}{\sqrt{1+\Delta}}\right),\sqrt{1+\Delta}\right)
			   +\frac{\Delta}{2} E_1\left(\pi/2-{\rm arccos}\left(\frac{\cos \alpha}{\sqrt{1+\Delta}}\right),\sqrt{1+\Delta}\right) & {\rm for}\qquad  \Delta<0,
			   \end{array}
			   \right.  \label{gl51}
\end{eqnarray}

$E_{2}$ being the elliptic integral of second kind. The surface part of the
excess free energy, from Eq.(\ref{gl41}),can be written as
\begin{equation}
F_{\rm surf}^{\rm bare} = -2h_1\cos \alpha + \frac{2}{\lambda} \cos^2 \alpha \label{gl52}
\end{equation}
From these results one can show immediately for $D \to \infty ,$
where $\Delta \to 0$  and $\phi_0 \to \pi/2$ hence $\alpha \to 0$, that
\begin{equation}
\Delta \approx 16 \exp(-D) - 8\alpha \exp(-D/2) \label{gl53}
\end{equation}
As anticipated above, $\Delta $ can be either positive ( if $ \alpha
<2\exp \left( -D/2\right)$ ) or negative ( if $\alpha >2\exp
\left( -D/2\right)$ ). In this limit the total free energy excess
that must be minimized becomes
\begin{equation}
F_{\rm exc}^{\rm total} \approx {\rm const} - 4\alpha \exp(-D/2) + \alpha^2\left( h_1-2/\lambda+1/2\right), \qquad \alpha<2\exp(-D/2) \label{gl53a}
\end{equation}
while for $\alpha>2\exp(-D/2)$ an extra term $4D\exp(-D/2)\left[\alpha-2\exp(-D/2)\right]$ has to be added in Eq.(\ref{gl53}). We are interested in the
wetting case $h_1>h_1^c$. In that case, minimization of $F_{\rm exc}^{\rm total}$
with respect to $\alpha$ yields
\begin{equation}
\alpha = 2\exp(-D/2)/\left( h_1-2/\lambda+1/2\right),  \label{gl54}
\end{equation}
which yields in Eq.(\ref{gl53})
\begin{equation}
\Delta = 16 \exp(-D) \frac{h_1-(1/2+2/\lambda)}{h_1+1/2-2/\lambda}. \label{gl55}
\end{equation}

Noting that $h_1^c = 1/2+2/\lambda$ and $w_{0}\left( \infty \right) =1$ in our units, the
final result for the reduction of the interfacial width can be cast in a
form very similar to Eq.(\ref{gl21}), namely
\begin{equation}
\frac{w_0(D)}{w_0(\infty)} = \left[ 1 + 16\frac{(h_1 - h_1^c)\exp(-D/w_0(\infty))}{1+h_1- h_1^c}
			     \right]^{-1/2}  \label{gl56}.
\end{equation}

Also for the opposite limit $D\to 0$, $\alpha \to \pi/2$ an
explicit analytical result is easily derived, since in this limit
\begin{equation}
D \approx 2 w_0 \left(\frac{\pi}{2}-\alpha \right) \label{gl57}
\end{equation}
and
\begin{equation}
F_{\rm exc}^{\rm tot} \approx \frac{D}{4} + \frac{D}{4w_0^2} - 2 h_1\cos \alpha + \frac{2}{\lambda} \cos^2 \alpha \approx \frac{D}{4} - 2h_1\left( \frac{\pi}{2}-\alpha \right)
+ \left( \frac{2}{\lambda} + \frac{1}{D}\right)\left( \frac{\pi}{2}-\alpha \right)^2. \label{gl58}
\end{equation}
Minimization with respect to $\alpha $ yields
\begin{equation}
\frac{\pi}{2} - \alpha \approx \frac{h_1}{2/\lambda+1/D} \approx h_1D, \qquad D \to 0  \label{gl59}
\end{equation}
and hence we obtain an equation analogous to the weak segregation limit,
\begin{equation}
w_0(D) \approx h_1/2. \label{gl60}
\end{equation}

\section{Numerical self-consistent field calculations for confined
interfaces of polymer mixtures with finite chain length}

We start the treatment by writing the partition function ${\cal Z}$ of a system of $%
n_{A}$ chains of type $A$ and $n_{B}$ chains of type $B$ in the volume $V$ with interactions ${\cal E}(\hat{\rho}_A,\hat{\rho}_B)$ as a
functional integral\cite{54}
\begin{equation}
{\cal Z} = \frac{1}{n_A!n_B!} \int {\cal D}[\vec{r}_\alpha] {\cal P}[\vec{r}_\alpha]
				   {\cal D}[\vec{r}_\beta ] {\cal P}[\vec{r}_\beta]
				   \exp \left[ - \frac{\Phi}{k_BT} \int {\rm d}^3\vec{r} \;\;\; {\cal E}(\hat{\rho}_A,\hat{\rho}_B)
				   \right] \label{gl61}
				   \end{equation}
where ${\cal D}[\vec{r}_\alpha]$ stands for the
functional integration of the coordinates of all the monomers of type $%
A$, ${\cal D}[\vec{r}_\beta ]$ the corresponding
term for the monomers of type $B$, and ${\cal P}[\vec{r}_\alpha]$, ${\cal P}[\vec{r}_\beta]$
are the corresponding probability distributions of the chain conformations in
the reference system. Since we compare the results of the SCF calculations with simulation data\cite{48} we use
the bond fluctuation model in the athermal limit as a reference system. In the framework of this
coarse grained lattice model each effective monomer blocks all 8 sites of an elementary cube of the lattice,
and working at $\Phi=1/16$ where half of the available sites are filled corresponds to a dense melt\cite{21}.
In the following all lengths are measured in units of the lattice spacing. 

Defining density operators as
\begin{equation}
\hat{\rho}_A(\vec{r}) = \frac{1}{\Phi} \sum_{\alpha=1}^{n_A} \sum_{i=1}^{N_A} \delta \left(\vec{r}-\vec{r}_{\alpha,i}\right) \label{gl62}
\end{equation}
where the first sum runs over all A chains and the second sum over all
monomers of the $\alpha $'th chain, being at positions $\vec{r}_{\alpha,i}$
, and similarly $\hat{\rho}_B(\vec{r})$. The normalized energy expression
can be written as
\begin{equation}
\frac{{\cal E}}{k_BT} = \frac{\zeta}{2} \left( \rho_A+\rho_B-1\right)^2 
- \frac{q\epsilon}{2} \left(\rho_A-\rho_B\right) \left[ 1+\frac{1}{2} l_0^2 \frac{\partial^2}{\partial z^2}\right]\left(\rho_A-\rho_B\right)
\label{gl63}
\end{equation}
where the first term is identical to the second term of Eq.(\ref{gl26}), while the
second term represents a pairwise monomer-monomer interaction of finite
range $l_0$ (in practice we take $l_0^2=16/9,$ when we try to
represent simulation results of the bond fluctuation model). Here $q$ is an
effective coordination number, and $\epsilon$ a normalized energy between a pair of
monomers ($2q\epsilon =\chi$, the Flory-Huggins interaction parameter)\cite{21}.
While in the previous section we have considered the limit where $%
N_{A}=N_{B}=N\to \infty $ and $\zeta \rightarrow \infty $
(incompressible melt of infinitely long chains), we here relax both these
approximations simultaneously, allowing both N finite (in the numerical
example shown in Fig. 3 we have chosen N = 32), and the compressibility $\zeta
$ also is taken finite ($\phi =1/16$ corresponds to $\zeta =4.1,$ as discussed
elsewhere \cite{21}).

Eqs.(\ref{gl61}) - (\ref{gl63}) is a very general formulation of the statistical mechanics of
polymers, which we here drastically simplify in terms of a mean field
approximation. The free energy density $f$ then can be expressed as $({\cal F} = - k_BT \ln {\cal Z})$.
\begin{eqnarray}
f=\frac{{\cal F}}{\Phi k_BT V} &=& \frac{\bar{\rho}_A}{N_A} \ln \bar{\rho}_A +  \frac{\bar{\rho}_B}{N_B} \ln \bar{\rho}_B +  \frac{1}{V} \int d^3r \; {\cal E}(\rho_A,\rho_B)  
    \nonumber \\
  && -  \frac{1}{V} \int {\rm d}^3\vec{r} \left\{  w_A\rho_A + w_B\rho_B \right\} -  \frac{\bar{\rho}_A}{N_A} \ln z_A[w_A] -  \frac{\bar{\rho}_B}{N_B} \ln z_B[w_B]
\label{gl64}
\end{eqnarray}
with
$\bar{\rho}_A = 1 - \bar{\rho}_B = \frac{n_AN_A}{\Phi V}$, $w_A$, $w_B$ being the effective fields, and $z_A$,$z_B$ the partition functions of single chains,
\begin{equation}
z_\alpha = \frac{1}{V} \int  {\cal D}[\vec{r}_\alpha] {\cal P}[\vec{r}_\alpha] \exp\left[-\sum_{i=1}^{N_\alpha} w_\alpha(\vec{r}_i) \right], \qquad \alpha=A,B \label{gl65}
\end{equation}
The self-consistent fields follow from:
\begin{equation}
w_A = \zeta \left(\rho_A+\rho_B-1\right) - q\epsilon\left[ 1+\frac{1}{2} l_0^2 \frac{\partial^2}{\partial z^2} \right] \left(\rho_A-\rho_B\right), \label{gl66}
\end{equation}
\begin{equation}
\rho_A = \frac{\bar{\rho_A}V}{N_Az_A} \frac{{\cal D}z_A}{{\cal D}w_A} = \bar{\rho}_A \sum_{\alpha} {\cal P}_w \frac{1}{N_A} \sum_{i=1}^{N_A} V \delta\left(\vec{r}-\vec{r}_{\alpha,i}\right)
\exp \left[ -\sum_{i=1}^{N_\alpha} w_\alpha(\vec{r}_{\alpha,i}) \right]/{\cal N} \label{gl67}
\end{equation}
where ${\cal N}$ is a normalizing denominator, ${\cal N} = \sum_{\alpha} {\cal P}_w  \exp \left[ -\sum_{i=1}^{N_\alpha} w_\alpha(\vec{r}_{\alpha,i}) \right]$,
the sum over $\alpha$ is a sum over a representative sample of (Monte-Carlo-generated) polymer conformations (in practice $7 \cdot 10^6$ $A$ chains and 
$7 \cdot 10^6$ $B$ chains are used), and the probability ${\cal P}_w$ of the confined polymer, in unnormalized form, is
\begin{equation}
{\cal P}_w = \left\{ \begin{array}{l}
	       0 \\
	       \exp(\pm \epsilon_w n_w)
		     \end{array}\right. \label{gl68}
\end{equation}
Here, ${\cal P}_w=0$ applies if any monomer falls outside of the walls of the film, while otherwise the weight depends on the number $n_w$ of monomers experiencing the
potential due to the wall $\epsilon_w$: the + sign applies for $A$ monomers at the left wall or $B$ monomers at the right wall, while the -- sign applies for $B$
monomers at the left wall or $A$ monomers at the right wall.

In practice the unknown functions $\rho_A(\vec{r})$, $\rho_B(\vec{r})$ that result from the solution of Eqs.(\ref{gl66}) and (\ref{gl67}) are found by choosing a Fourier decomposition
into a set of basis functions appropriate for the chosen geometry,
\begin{equation}
f_k(z) = \sqrt{2} \sin \frac{\pi k z}{D} , \qquad k=1,2,\cdots; \label{gl69}
\end{equation}
The coefficients of this Fourier decomposition are found iteratively by the Newton-Raphson method. Note that the most difficult part of the present numerical SCF
scheme actually is the summation over the sample of polymer conformations, which is done on a CRAY T3E multiprocessor machine.

Fig.\ 4 shows typical profiles resulting from this method for $N=32$ and three choices of $\epsilon$. While for $\epsilon=0.03$ we have a strongly segregated case
quite comparable to the corresponding Monte Carlo results, for $\epsilon=0.016$ we have a profile that already develops the shape characteristic of the weak segregation case (Fig.\ 1).
Fig.\ 5 shows the thickness dependence of $w(D)$ resulting for this model at $\epsilon=0.03$, comparing the SCF results with the corresponding Monte Carlo results from
Werner {\em et al}\cite{48}. Note that these Monte Carlo results are not the full width of the profiles shown in Fig.\ 3 -- which include also effects due to capillary
wave broadening that cannot be taken into account by the above SCF treatment -- but rather are constraint by recording the mean square interfacial width only on a lateral length scale $B=8$ lattice spacings (the system is divided into a grid of $B \times B$ subblocks, and the local position of the interface in each subblock is recorded
separately to obtain this local width). The proper choice of the value of this grid size $B$ is discussed elsewhere\cite{48,49}.

\section{Discussion and conclusions}
In this paper we have considered the change of the ``intrinsic'' width $w_0(D)$ of an interface between coexisting phases confined in a thin film
between ``competing walls'' a distance $D$ apart. We have considered first two limiting cases which can both be treated by ``square gradient''-type theories,
namely the Cahn-Hilliard theory of a weakly segregated symmetric binary mixture (Sec.II) and the Helfand theory of a strongly segregated incompatible polymer mixture in the limit of infinite chain lengths (Sec.III). In both cases the reduction of the interfacial width $w_o(D)$ due to confinement can be easily worked out, and a region where $w_0(D)$ varies approximately linearly with $D$ occurs in
both cases. Treating finite chain lengths for polymers, a numerical scheme has been used (Sec.IV) which also allows the treatment of cases intermediate between weak and strong segregation (Fig.\ 4).
In this way, the different regimes proposed for the concentration profile in Fig.\ 1 quantitatively could be demonstrated explicitly.

There is one important drawback of our treatment, however: while all variants of mean field theories (like those presented in the previous sections) readily yield ``intrinsic'' interfacial profiles, the latter are \underline{not} well-defined in the framework of rigorous statistical mechanics, and consequently there is no unique way to define them either in a computer simulation
nor in an experiment on real materials. The actual interfacial width $w(D)$ observed in both simulations and experiments exhibits an additional broadening due to fluctuations in the local position of the 
center $h(x,y)$ of the interface away from its average position $\langle h(x,y) \rangle$ ($=0$, in our choice of coordinate system, where the plane $z=0$ of the $(x,y,z)$-coordinate system is halfway in 
between the confining walls).

Werner {\em et al.}\cite{48} have discussed the extent to which one can describe this broadening in terms of a convolution approximation,
\begin{equation}
\rho_A(z) = \int_{-D/2}^{+D/2} {\rm d}h\; \rho_A^{\rm int}(z-h)P_D(h), \label{gl71}
\end{equation}
where $ \rho_A^{\rm int}(z-h)$ is the intrinsic profile, for an interface centered at $z=h$, and $P_D(h)$ describes the probability that a deviation $h$ from the average value $\langle h \rangle = 0$
occurs for a film thickness $D$. It then was assumed that this probability distribution is a Gaussian, $P_D(h) = \exp(-h^2/2s^2)/\sqrt{2\pi s^2}$, and hence one finds for the total mean square width
\begin{equation}
w^2(D) = w_0^2(D) + \frac{\pi}{2} s^2(D) \label{gl72}
\end{equation}

In order to calculate the additional broadening due to the interfacial position fluctuations, $s^2(D)$, an approximation was used where the interface is described by an effective interface Hamiltonian
describing capillary waves in a harmonic potential:
\begin{equation}
{\cal H}_{\rm eff}(h) = \int {\rm d}x{\rm d}y\;\; \left\{  \frac{\sigma(D)}{2} \left[\left(\frac{\partial h}{\partial x}\right)^2 + \left(\frac{\partial h}{\partial y}\right)^2 \right]  
+\frac{a}{2} \exp\left( -\frac{\kappa D}{2}\right)h^2\right\}  \label{gl73}
\end{equation}
where $\sigma(D)$ is an effective interfacial stiffness (which converges to the interfacial stiffness $\sigma \equiv \sigma(\infty)$ of a free unconfined interfaces as $D \to \infty$), $a$ is a constant, and $\kappa^{-1}$ is a decay length which is of the order of the correlation length $\xi_b$ in the weak segregation case, but of the order of $w_0(\infty) = b/\sqrt{6\chi}$ in the limiting case
of strongly segregated polymer mixture with $N \to \infty$. Using Eq.(\ref{gl73}), one then obtains that $s^2(D)\propto D$ for large $D$, since
\begin{equation}
s^2(D) = \frac{1}{2\pi} \int_0^{2\pi/B} \frac{{\rm d}q}{\sigma(D)q^2+a\exp(-\kappa D/2)} \to \frac{\kappa (D)D}{8\pi \sigma(D)} + {\rm const} \label{gl74}
\end{equation}
Note that short wavelengths need to be cut off if they are less than some length $B$, in order that the integral in Eq.(\ref{gl74}) converges. For self-consistency, all fluctuations on length scales
smaller than $B$ must be included in the intrinsic width $w_0(D)$. However, there is no obvious theoretical recipe that would uniquely define this cutoff $B$, and thus the separation of
interfacial fluctuation into ``intrinsic'' and capillary wave''-type is somewhat arbitrary. In the simulations, $B=8$ lattice spacings was chosen simply for the reason that then $w_0(D\to\infty)$
agrees with the self-consistent filed calculations, and as Fig.\ 5 demonstrates, there is then fair agreement between the SCF prediction for $w_o(D)$ and the Monte Carlo observation for all $D$.
However, apart from this fact one has no reason for not chosing $B=7$ or $B=9$, for instance. Also the analysis of the capillary wave spectrum in the confined geometry did reveal a pronounced dependence of $\sigma(D)$ on $D$
in the same regime where $w_0(D)$ differs appreciably from $w_0(\infty)$. As yet, an analytical approach to accurately predict the interface stiffening ($\sigma(D)$ is enhanced for small $D$) due to
confinement is lacking. It is conceivable that also the constants $a$ and $\kappa$ of the effective interface potential $V_D(h) = a\exp(-\kappa D/2) h^2/2$ are no true constants but also depend
weakly on $D$.

Particularly cumbersome is the theoretical understanding of the crossover between the weak segregation case and the strong segregation limit of a polymer mixture. Just as two lengths control the 
interfacial profile\cite{12}, the length $w_0(\infty)=b/\sqrt{6\chi}$ shows up in the center of the profile, the radius of gyration $R_g = b\sqrt{N/6}$ in the wings, we expect two decay constants 
in the potential $V_D(h)$,namely
\begin{equation}
V_D(h) = \frac{1}{2} h^2 \left\{ a \exp(-D/w_0(\infty)) + a' \exp(-D/2R_g) \right\}  \label{gl75}
\end{equation}
While the amplitude $a'$ of the second term vanishes in the strong segregation limit, when $m_b \to 1$, it decays much slower with increasing $D$ than the first term, for large $N$, and, hence
the interplay between these terms is subtle. We expect that a similar expression will interpolate between our weak segregation result for $w_0(D)$ (Eq.{\ref{gl21})),
and the strong segregation result Eq.(\ref{gl57}).
Similarly,it is not clear whether the cutoff length $B$ should be of the order of $w_0(\infty)$ or of the order of $R_g$, in this case. Finally, we remind
the reader that the exponential variation of $V_D(h)$ with $D$ is only appropriate for short range forces between the walls and the molecules of the mixture, not for the --
physically more realistic -- long range van der Waals forces. Thus, our treatment is a first step towards the resolution of a rather complex problem only. But there is clear 
experimental evidence\cite{47,50} that the effects discussed here are indeed practically relevant. Thus we hope that our study will stimulate further efforts to understanding this problem.

\subsection*{Acknowledgments}
Support by the Deutsche Forschungsgemeinschaft (DFG), Grant No Bi314/18, and by the Bundesministerium f\"ur Bildung, Wissenschaft, Forschung und Technologie (BMBF), Grant No 03N8008C, is
gratefully acknowledged. The authors thank J. Klein and T. Kerle for stimulating discussions.

\newpage
\begin{figure}[htbp]
\begin{minipage}[t]{160mm}%
   \mbox{
   \hspace*{-2cm}
   \setlength{\epsfxsize}{14cm}
   \epsffile{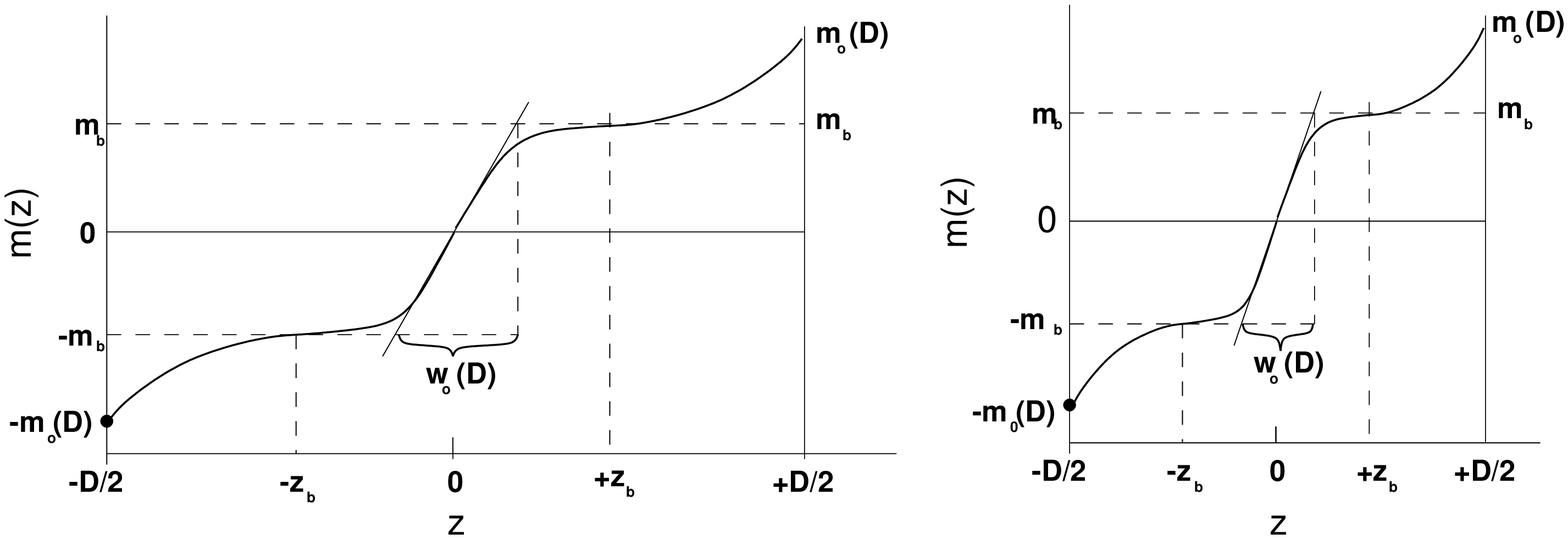}
   }
\end{minipage}%
\hfill
\begin{minipage}[b]{160mm}%
\caption{
\label{fig:1}
}
Schematic order parameter profiles $m(z)$ versus $z$, for a symmetrical phase separated mixture confined between two competing walls, such
that the left wall prefers the $B$-rich phase and the right wall prefers the $A$-rich phase. The upper part shows a large thickness $D$
of the thin film, the lower part a smaller thickness. The definition of the intrinsic thickness-dependent interfacial width $w_0(D)$ is indicated.
For further explanations cf.\ text.
\end{minipage}%
\end{figure}

\begin{figure}[htbp]
\begin{minipage}[t]{160mm}%
   \mbox{
   \hspace*{-2cm}
   \mbox{
     \setlength{\epsfxsize}{7cm}
     \epsffile{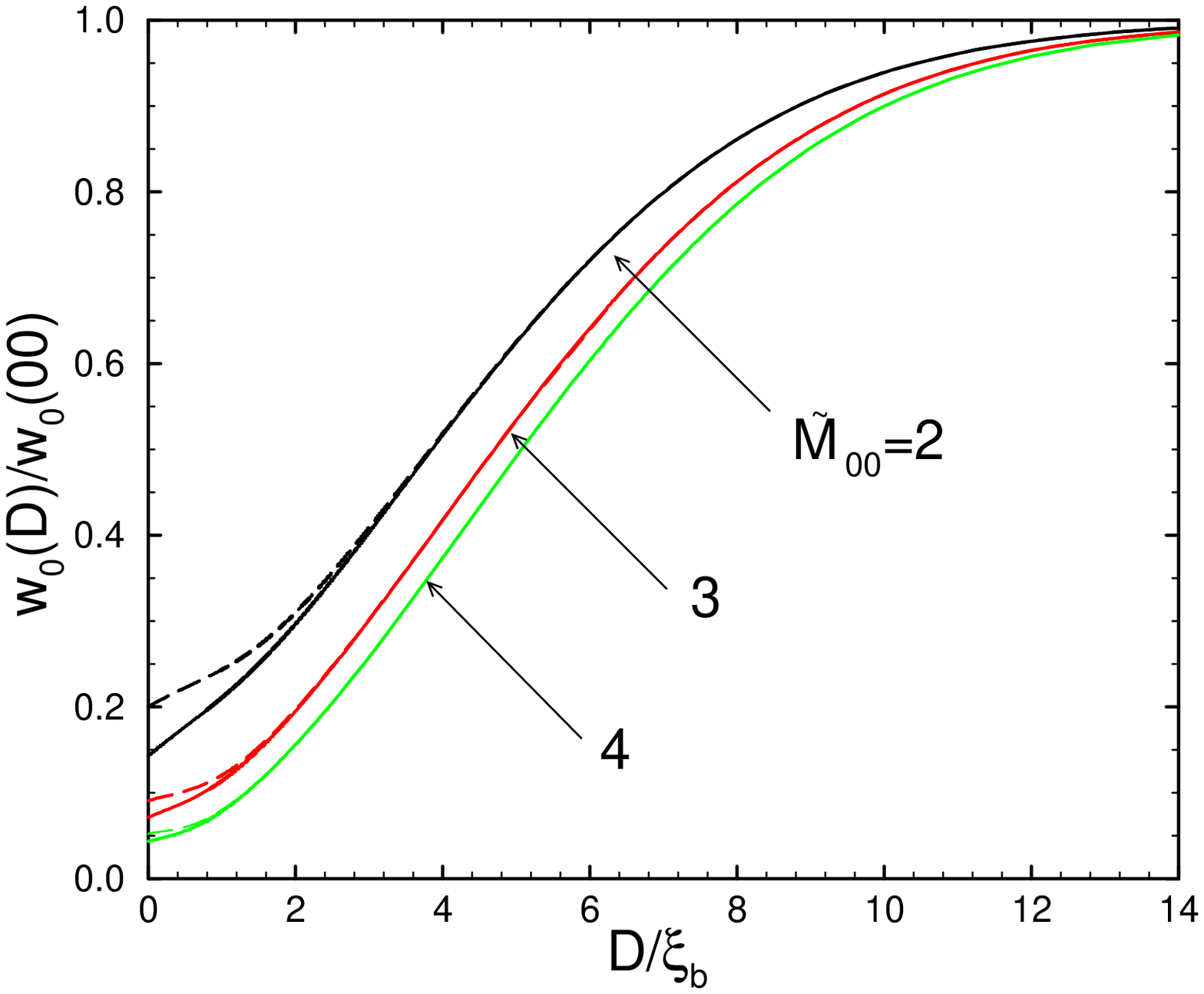}
     \setlength{\epsfxsize}{7cm}
     \epsffile{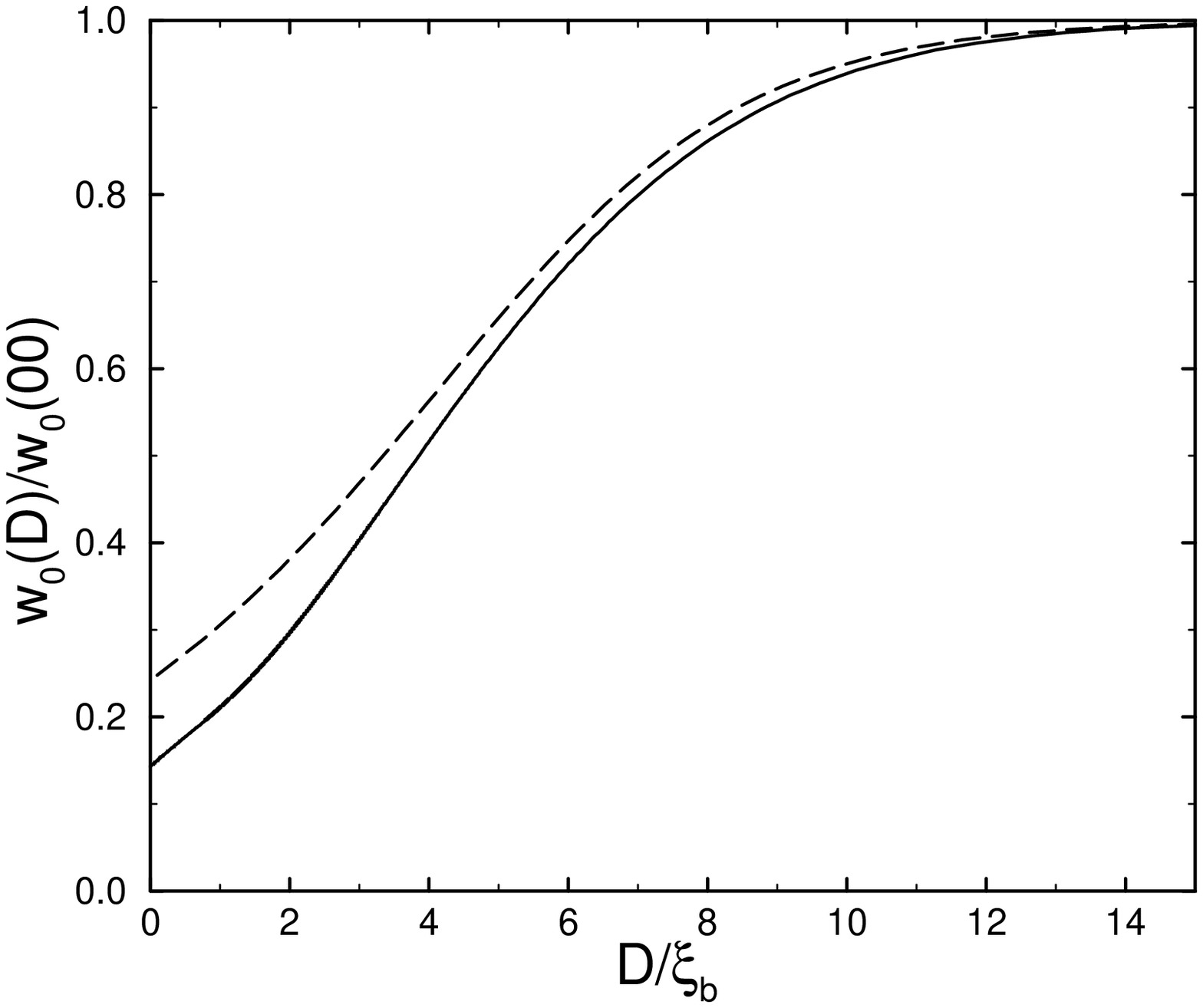}
   }
   }
\end{minipage}%
\hfill
\begin{minipage}[b]{160mm}%
\caption{
\label{fig:2}
}
{\bf a}) Plot of the reduced interfacial widths $w_0(D)/w_0(\infty)$ versus $D/\xi_b$ for $\tilde{M}_\infty =$ 2,3,and 4,
and two choices of $\lambda/\xi_b$:  $\lambda/\xi_b=1$ (full curves) and $\lambda/\xi_b=2$ (broken curves). \\
{\bf b}) Comparison of the exact numerical result for $w_0(D)/w_0(\infty)$ versus $D/\xi_b$ for the choice $\tilde{M}_\infty =$ 2,
$\lambda/\xi_b=1$ with the approximation, Eq.(\ref{gl20}) (broken curve).
\end{minipage}%
\end{figure}

\begin{figure}[htbp]
\begin{minipage}[t]{160mm}%
   \hspace*{-2cm}
   \mbox{
     \setlength{\epsfxsize}{7cm}
     \epsffile{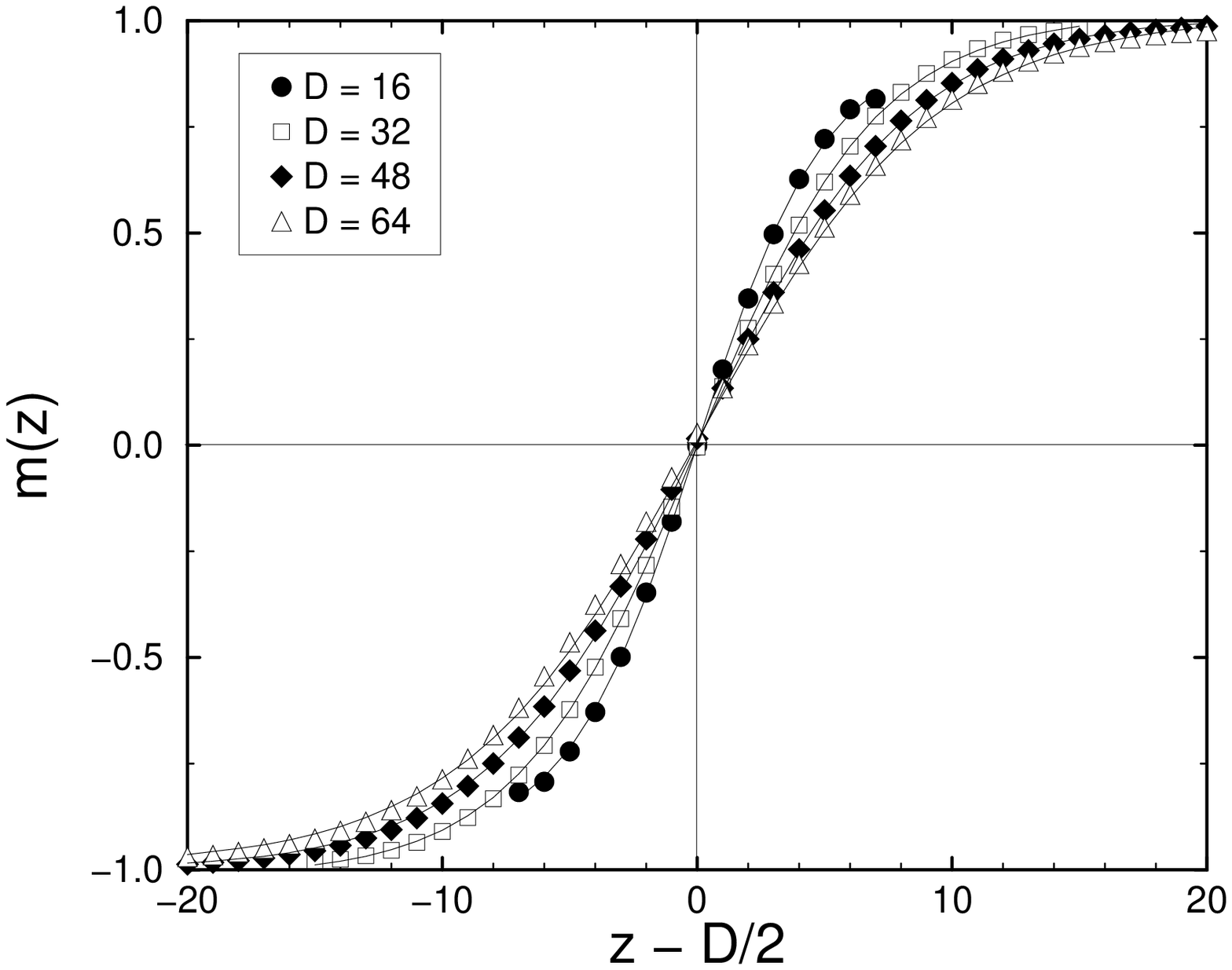}
     \setlength{\epsfxsize}{7cm}
     \epsffile{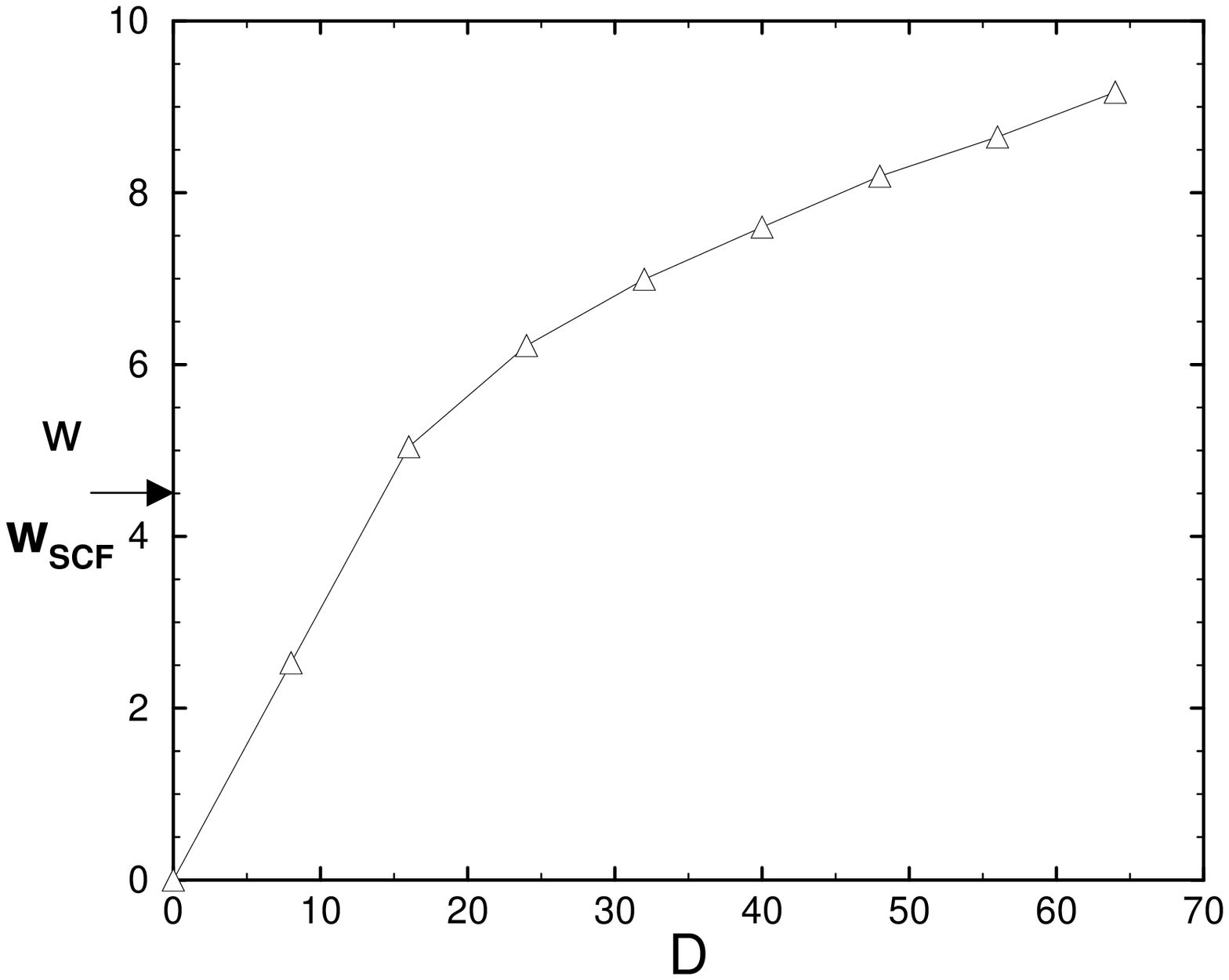}
   }
\end{minipage}%
\hfill
\begin{minipage}[b]{160mm}%
\caption{
\label{fig:3}
}
{\bf a})
Order parameter profiles $m(z)$ versus $z$ for films of thickness $D=16,32,48$, and 64 for the bond fluctuation model
of a symmetrical polymer mixture ($N_A=N_B=N=32$, $\epsilon_{AA}=\epsilon_{BB}=-\epsilon_{AB}=-k_BT\epsilon$ with $\epsilon=0.03$
corresponding to $T/T_{cb}=0.48$,$\epsilon_w=\pm0.1$ being a wall-monomer interaction of square well type and a range of 2 lattice spacings),
using a $L \times L \times D$ geometry with two $L \times L$ surfaces along which periodic boundary conditions act.
All lengths are measured in units of the lattice spacing. 
{\bf b}) Interfacial width $w$ plotted versus $D$.
Note that for $D<20$ we have $w \propto D$ here, while the value $w_0(\infty)$ predicted by the self-consistent field theory (SCF) is shown as an arrow on the ordinate. 
From Werner {\em et al}\cite{48}.\\
\end{minipage}%
\end{figure}

\begin{figure}[htbp]
\begin{minipage}[t]{160mm}%
   \mbox{
   \hspace*{-2cm}
   \mbox{
     \setlength{\epsfxsize}{9cm}
     \epsffile{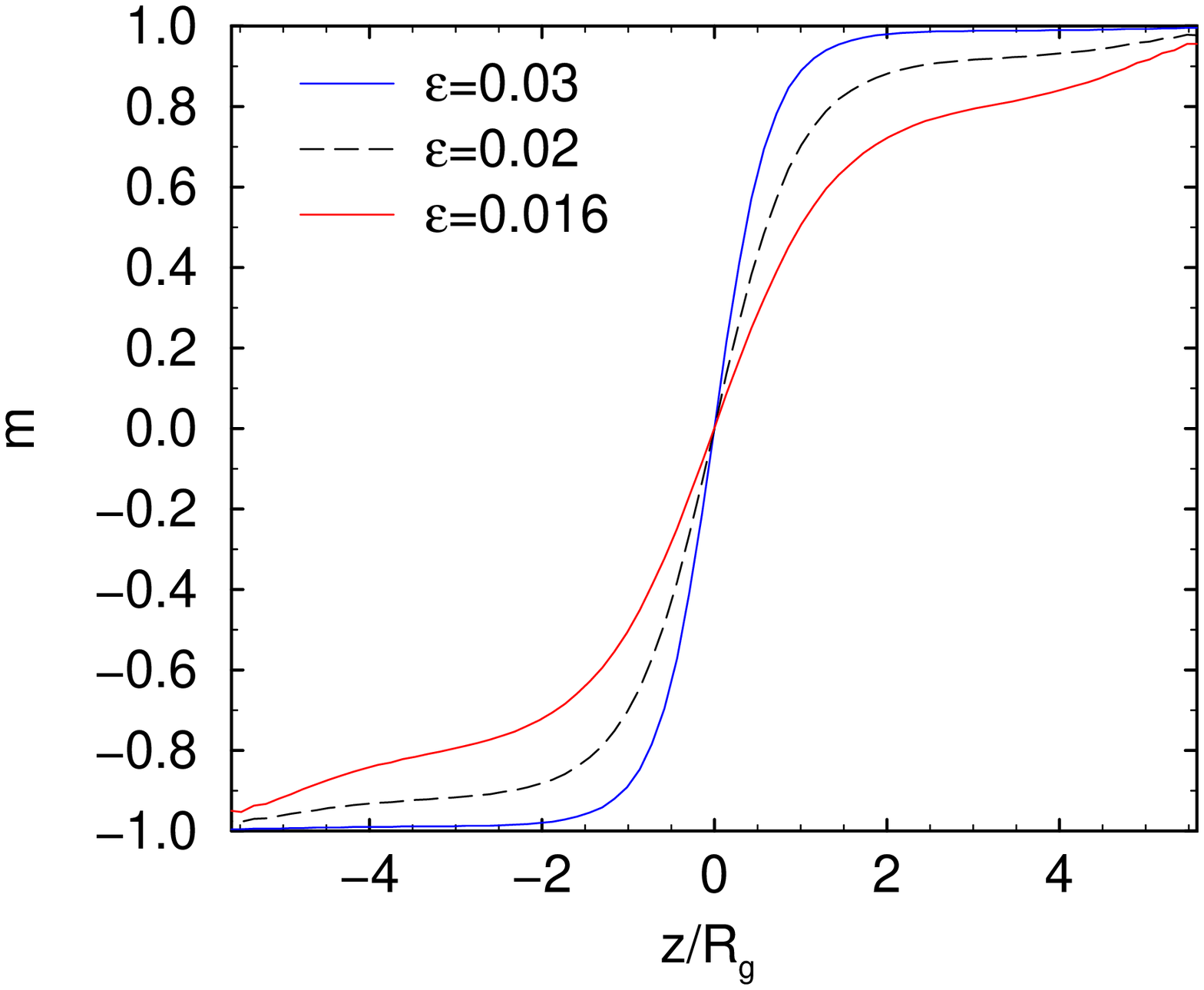}
   }
   }
\end{minipage}%
\hfill
\begin{minipage}[b]{160mm}%
\caption{
\label{fig:4}
}
Plot of the intrinsic order parameter profile $m(z)$ {\em vs.} $z$, in units of the radius of gyration  for chain length $N=32$ and three
choices of $\epsilon$. Curves result from the numerical self-consistent field scheme, Eqs.(\ref{gl62})-(\ref{gl69}).
\end{minipage}%
\end{figure}

\begin{figure}[htbp]
\begin{minipage}[t]{160mm}%
   \mbox{
   \hspace*{-2cm}
   \mbox{
     \setlength{\epsfxsize}{9cm}
     \epsffile{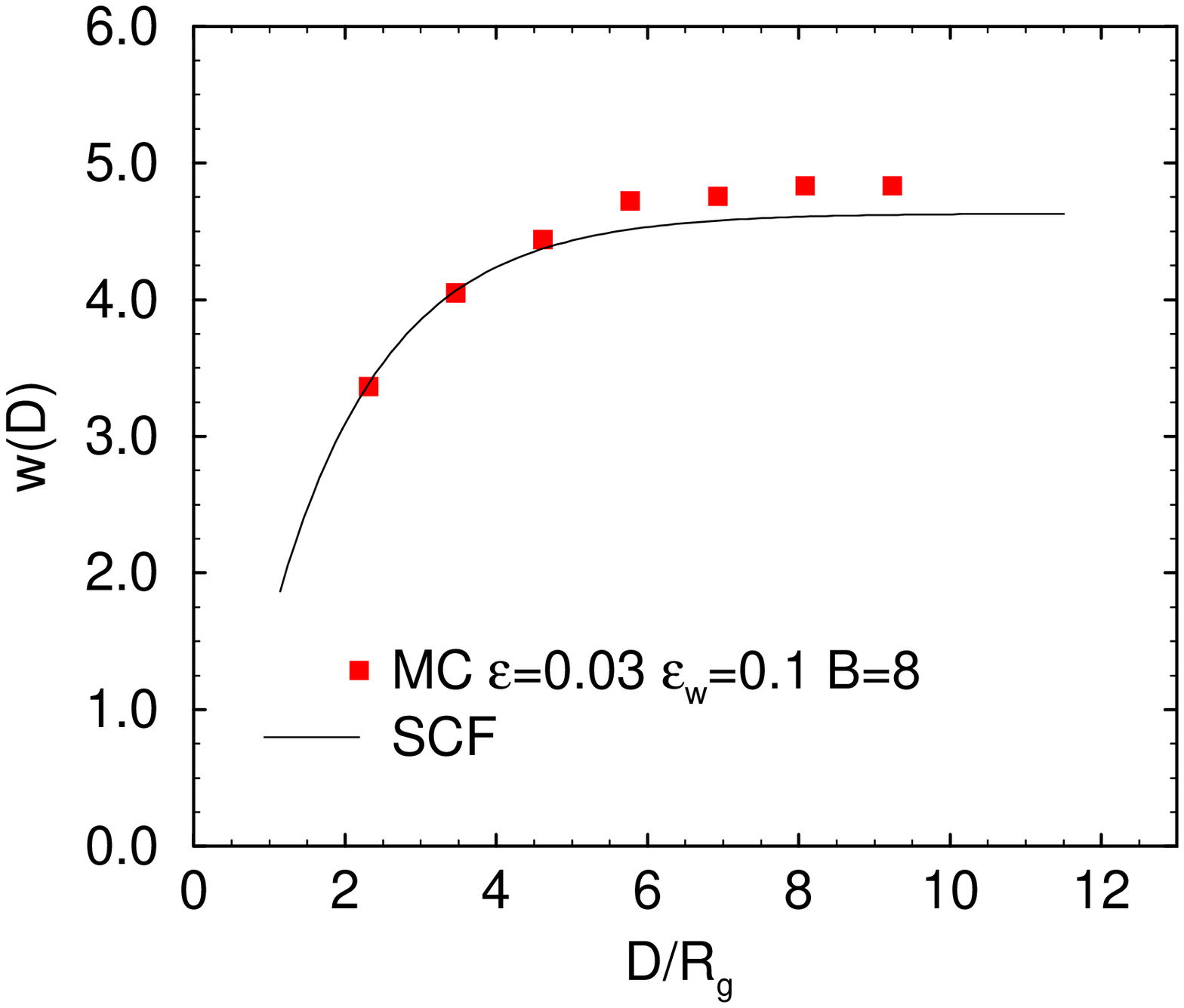}
   }
   }
\end{minipage}%
\hfill
\begin{minipage}[b]{160mm}%
\caption{
\label{fig:5}
}
Intrinsic width $w(D)$ plotted {\em vs} $D/R_g$ for polymers of chain length $N=32$, and energy parameters $\epsilon=0.03$, $\epsilon_w=0.1$.
The curve shows the result extracted from the numerical self-consistent field scheme $\{$ Eqs.(\ref{gl62})-(70)$\}$,
while squares are the Monte Carlo data of Werner {\em et al.}\cite{48}. For extracting the intrinsic width from the simulations a lateral grid size 
$B \times B$ (with $B=8$ lattice spacings) was used\cite{48}.
\end{minipage}%
\end{figure}

\end{document}